\documentstyle[12pt]{article}

\newcommand{\bq}{\begin{equation}}
\newcommand{\eq}{\end{equation}}
\newcommand{\bqn}{\begin{eqnarray}}
\newcommand{\eqn}{\end{eqnarray}}
\newcommand{\nb}{\nonumber}
\newcommand{\lb}{\label} 

\begin{document} 
\baselineskip 0.86cm
\title{Instability of cosmological event 
horizons of non-static global
cosmic strings}
 \author{ 
Anzhong Wang \thanks{e-mail address: wang@symbcomp.uerj.br}\\
\small Departamento de F\' {\i}sica Te\' orica,
Universidade do Estado do Rio de Janeiro, \\
\small Rua S\~ ao Francisco Xavier 524, Maracan\~ a,
20550-013 Rio de Janeiro~--~RJ, Brazil\\
\small and\\
\small Observat\'orio Nacional~--~CNPq, Rua General Jos\'e Cristino 77,\\
\small S\~ao Crist\'ov\~ao, 20921-400 Rio de Janeiro~--~RJ, Brazil\\
Jos\'e A.C. Nogales \thanks{e-mail:jnogales@portela.if.uff.br}\\
\small Instituto de F\' {\i}sica, Universidade Federal Fluminense,\\
\small Av. Litor\^anea s$/$n - Boa Viagen,  CEP 24210-340  Niter\'oi~--~RJ, 
Brazil\\
\small and\\
\small Instituto de F\' {\i}sica, Universidad Mayor de San Andres (UMSA) , \\ 
\small Casilla 3553,  La Paz~--~Bolivia}
 
\date{}
\maketitle
\input epsf

\newpage

\begin{abstract}

\baselineskip 0.85cm

The stability of the cosmological event horizons found recently by
Gregory [Phys. Rev. D54, 4955 (1996)] for a class of non-static global
cosmic strings is studied. It is shown that they are not stable to both
test particles and physical perturbations. In particular, the back reaction of
the perturbations of null dust fluids will turn them into
spacetime singularities. The resulted singularities are strong in the
sense that the distortion of test particles diverges logarithmically
when these singular hypersurfaces are approaching.

\end{abstract}

\vspace{.4cm}


\newpage

\baselineskip 0.9cm

\section*{I. INTRODUCTION}

Topological defects formed in the early Universe have been studied
exclusively \cite{VS1994}, since the pioneering work of Kibble
\cite{Kibble1976}.  They were formed during phase transitions of the
Universe, where the degenerated vacua acquired non-zero expectation
values. Depending on the topology of the vacua, the defects could be
domain walls, cosmic strings, monopoles, textures, or the hybrids of
them. Among these defects, cosmic strings have received particular
attention mainly because of their cosmological implications: They might
provide the seeds for the formations of galaxies and the large-scale
structure of the Universe \cite{VS1994}.

Cosmic strings are further classified as local (gauge) and global
strings, according to whether they arise from a local symmetry breaking
or a global symmetry breaking. These two kinds of strings have very
different properties. In particular, the spacetime of a local (static)
string is well behaved and asymptotically approaches a conical spacetime
\cite{Gregory1987}, while the spacetime of a global static string is
necessarily singular at a finite distance from the symmetric axis, and
its deficit angle diverges logarithmically \cite{HS1988}.  It is this
undesirable property that make global strings very difficult to use,
and most studies of cosmic strings have been restricted only to local
strings \cite{VS1994}.

However, local strings are tightly constrained by their contribution to
the gravitational radiation background \cite{Bennett1991,WS1996}, while
global cosmic strings circumvent this constraint and may have similar
cosmological implications \cite{Battye1996}. Lately, Banerjee {\em et al.}
\cite{Banerjee1996} and Gregory \cite{Gregory1996} studied non-static
global strings, and some interested results were found. In particular,
Gregory showed that the spacetime singularities usually appearing in the
static case can be replaced by cosmological event horizons (CEH)
\footnote{Note that Gregory called the horizons as event horizons.
However, to be distinguishable with the ones of black holes, following
Gibbons and Hawking \cite{GH1977}, we call them cosmological event
horizons.}. This result is very important, as it may make the structure
formation scenario of cosmic strings more likely, and may open a new
avenue to the study of global strings.

In this paper, we shall study the stability of the CEHs found above by
Gregory \cite{Gregory1996}, and shall show that in general they are not
stable against small perturbations, instead are turned into spacetime
singularities. Does this mean that the hope that time-dependence might
remove the singular nature of global static string spacetimes is already found
negative? We think that it is not, at least as far as the work of
Gregory is concerned. As a matter of fact, Gregory considered a very
particular case: the energy-momentum tensor of the string is still
time-independent. As a result, no gravitational and particle radiation
exists. For more general case, one would expect that CEHs may not be
formed at all, or even they are formed but stable. As we know, in the
cylindrical case gravitational and particle radiation in general always
exists. It is plausible to expect that in some situations the
radiation is so strong that the gravitational field is well dilated
before any spacetime singularity or horizon is formed. 

The rest of the paper is organized as follows: In Section II we shall
briefly review the main properties of the spacetimes studied by Gregory
\cite{Gregory1996}, while in Section III null dust fluids of test
particles are studied, which indicate some singularity behavior of the
spacetimes near the CEHs. In Section IV we consider ``physical"
perturbations of real particles, and confirm the results obtained
in Section III. Here ``physical" is in the sense that the back reaction
of the perturbations are taken into account. Finally in Section V we
derive our main conclusions.

\section*{II. SPACETIME FOR NON-STATIC GLOBAL COSMIC STRINGS}

\renewcommand{\theequation}{2.\arabic{equation}}
\setcounter{equation}{0}

For a straight cosmic string, we can always choose a coordinate system
that is comoving with the string so that the spacetime in this system
has a cylindrical symmetry. If additionally we require that the string
has no rotation, the metric for such a spacetime takes the general form
\cite{MC1980} 
\bq
\lb{2.1}
ds^{2} = e^{2(\gamma - \psi)}(dt^{2} - dR^{2}) - e^{2\psi}dz^{2} 
- \alpha^{2}e^{- 2\psi}d\theta^{2},  
\eq
where $\gamma, \psi$ and $\alpha$ are functions of $t$ and $R$ only, and
$t, R, z$ and $\theta$ are the usual cylindrical coordinates.

By requiring that the string has fixed proper width and that the
spacetime has boost symmetry in the ($t, z$)-plane, Gregory managed to
show that the spacetime for a U(1) global string (vortex) is given by
\cite{Gregory1996} 
\bq
\lb{2.2}
\gamma = 2a(R) + b(t), \;\;\; \psi = a(R) + b(t),\;\;\; 
\alpha = c(R)e^{a(R) + b(t)},
\eq
where $a(R)$ and $c(R)$ are two arbitrary functions, and $b(t)$
is given by 
\bq
\lb{2.3}
b(t) = \left\{ \begin{array}{ll}
\ln[\cosh(\beta t)], \pm \beta t, & b_{0} > 0,\\
b_{1}\ln t, & b_{0} = 0,\\
\ln[cos(\beta t)], & b_{0} < 0,
\end{array} \right.
\eq
where $b_{0}$ and $b_{1}$ are arbitrary constants, and $\beta \equiv
\sqrt{|b_{0}|}$.  from Eqs.(\ref{2.1}) and (\ref{2.2}) we can see
that, by introducing a new radial coordinate $r$ via the relation
\bq
\lb{2.4}
r = \int^{r}{e^{a(R)}dR},
\eq
the metric (\ref{2.1}) can be written in the form 
\bq
\lb{2.5}
ds^{2} = e^{2 A(r)}dt^{2} - dr^{2} - e^{2[A(r) + b(t)]}dz^{2} 
- C^{2}(r)d\theta^{2},  
\eq
where $A(r) \equiv a(R(r))$ and $C(r) \equiv c(R(r))$. This is exactly
the form used by Gregory \cite{Gregory1996}.

As shown by Gregory, the spacetime inside the core of a string is always
singular at a finite distance for the cases $b_{0} \le 0$, and has a CEH
for $b_{0} > 0$. In fact, in the latter case the metric coefficients
have the asymptotic behavior
\bq
\lb{2.9}
e^{A(r)} \sim \beta(r_{0} - r),\;\;\; C(r) \sim C_{0} + O(r_{0} -
r)^{2}, \eq as $r \rightarrow r_{0}^{-}$, where $C_{0}$ is a constant
[cf. Eq.(3.14) in Ref. 9]. For the choice $b(t) = \ln[\cosh(\beta t)]$,
the corresponding metric takes the form
\bq
\lb{2.10}
ds^{2} = \beta^{2}(r_{0} - r)^{2}\left[dt^{2} - 
\cosh^{2}(\beta t) dz^{2}\right] - dr^{2}
- C_{0}^{2}d\theta^{2},  
\eq
in the neighborhood of the the hypersurface $r = r_{0}$. One can show
that the singularity appearing at $r = r_{0}$ in Eq.(\ref{2.10}) is a
coordinate one. This can be seen, for example, by making the following
coordinate transformations
\bqn
\lb{2.11}
X &=& (r_{0} - r)\cosh(\beta t)\cos(\beta z),\nb\\
Y &=& (r_{0} - r)\cosh(\beta t)\sin(\beta z),\nb\\
T &=& (r_{0} - r)\sinh(\beta t),
\eqn
then the metric (\ref{2.10}) is brought to the form
\bq
\lb{2.12}
ds^{2} =  dT^{2} - dX^{2} - dY^{2} - C_{0}^{2}d\theta^{2},  
\eq
which is locally Minkowski. Thus, the singularity at $r = r_{0}$ in the
coordinates $\{t, r, z, \theta\}$ is indeed a coordinate singularity and
represent a cone-like CEH in the coordinates $\{T, X, Y, \theta\}$, as
one can see from Eq.(\ref{2.11}) that the hypersurface $r = r_{0}$ is
mapped to
\bq
\lb{2.13}
X^{2} + Y^{2} = T^{2}, \; (r = r_{0}).
\eq
For the details, we refer readers to \cite{Gregory1996}.

For $b(t) = \pm \beta t$, the corresponding metric takes the
form
\bq
\lb{2.14}
ds^{2} = \beta^{2}(r_{0} - r)^{2}\left[dt^{2} - 
e^{\pm \beta t} dz^{2}\right] - dr^{2}
- C_{0}^{2}d\theta^{2}.  
\eq
One can show that the singularities appearing at $r = r_{0}$ in the
above metric also represent CEHs. In fact, if we make the
coordinate transformations 
\bqn
\lb{2.15}
T &=& \frac{1}{2}(r_{0} - r)\left[\beta^{2}z^{2}e^{\beta t} + 
2 \sinh(\beta t)\right],\nb\\
X &=& \frac{1}{2}(r_{0} - r)\left[\beta^{2}z^{2}e^{\beta t} - 
2 \cosh(\beta t)\right],\nb\\
Y &=& \beta (r_{0} - r)ze^{\beta t},  
\eqn
for $b(t) = + \beta t$, the metric (\ref{2.14}) will be brought to the
exact form of Eq.(\ref{2.12}), while if we make the same transformations
as those of Eq.(\ref{2.15}) but with $t$ being replaced by $- t$ for
$b(t) = - \beta t$, the metric (\ref{2.14}) will be also brought to the
same form, Eq.(\ref{2.12}). Therefore, in the latter two  cases the
hypersurface $r = r_{0}$ all represents a CEH. The topology of it is
also conical in the Minkowiski-like coordinates ($T, X, Y, \theta$), 
as one can
show from Eq.(\ref{2.15}) that Eq.(\ref{2.13}) is satisfied, too.

Before proceeding further, we would like to note the following: a) For
the coordinate transformations given by Eq.(\ref{2.11}), the mapping
between ($t, r, z$) and ($T, X, Y$) is not one-to-one, while the one
given by Eq.(\ref{2.15}) is. b) The nature of the
CEHs in all these three cases is quite similar to that of the extreme
Reissner-Nordstr\"om black hole \cite{HE1973}, in the sense that across
$r = r_{0}$ the coordinate $t$ remains time-like, while $r$
remains space-like. c) in the neighborhood $r = r_{0}$ but with $r >
r_{0}$, Eq.(\ref{2.9}) should be replaced by
\bq
\lb{2.17}
e^{A(r)} \sim \beta(r - r_{0}),\;\;\; C(r) \sim C_{0} +
 O(r - r_{0})^{2}, \;(r > r_{0}).
\eq
Substituting Eqs.(\ref{2.9}) and (\ref{2.17}) into Eq.(\ref{2.4}),
we find
\bq
\lb{2.18}
R = \left\{ \begin{array}{ll}
\frac{1}{\beta}\ln[\beta(r - r_{0})], & r > r_{0},\\
- \frac{1}{\beta}\ln[\beta(r_{0} - r)], & r < r_{0},
\end{array} \right.
\eq
which shows that $R$ is a monotonically increasing function of $r$, except
for the point $r = r_{0}$, at which $R$ diverges.  
 
In the following, we shall consider the stability of these CEHs in two
steps: First, in the next section we shall consider test particles near
the CEHs along a line suggested by Helliwell and Konkowski (HK)  in the
study of the stability of quasiregular singularities \cite{HK1985}, at
the aim of generalizing the HK conjecture to the case of CEHs. Second,
in Sec. IV we shall consider perturbations of null dust fluids. These
perturbations are different from the ones studied in Sec. III, in the
sense that the back reaction of them to the spacetime backgrounds will
be taken into account. 

\section*{III. TEST NULL DUST FIELDS NEAR THE CEHs}

\renewcommand{\theequation}{3.\arabic{equation}}
\setcounter{equation}{0}

In a series of papers \cite{HK1985}, HK studied the stability of
quasiregular singularities by using test fields. In particular, they
conjectured that {\em if one introduces a test field whose energy-momentum
tensor (EMT) calculated in a freely-falling frame mimics the behavior of
the Riemann tensor components that indicate a particular type of
singularity (quasiregular, non-scalar curvature, or scalar curvature),
then a complete non-linear back-reaction calculation would show that
this type of singularity actually occurs}. Recently, this conjecture was
further generalized to the stability of Cauchy horizons \cite{HK1992}.
Clearly, if this conjecture is true, the stability analysis of spacetime
singularities would be considerably simplified.  In this section, using
HK's ideas we shall study test null dust fields near the CEHs.

For a null dust fluid moving along the outgoing null geodesics defined
by $n^{\mu}$ in the region $r \le r_{0}$, the EMT takes the form
\bq
\lb{3.1}
T^{- out}_{\mu\nu} = \rho^{out}_{-}  n_{\mu} n_{\nu},
\eq
where $n_{\mu}$ is a null vector defined as that in Eq.(\ref{A.11}).
Then, from the conservation equations $ T^{- out}_{\mu\nu;\lambda}
g^{\nu\lambda} = 0$, we find
\bq
\lb{3.2}
e^{A} \left(\frac{\rho^{out}_{-} ,_{r}}{\rho^{out}_{-} } +
3A'(r)\right) +  \left(\frac{\rho^{out}_{-} ,_{t}}{\rho^{out}_{-} } +
b'(t)\right) = 0,
\eq
which has the solution
\bq
\lb{3.3}
 \rho^{out}_{-}  = \frac{\rho^{out (0)}_{-} e^{a_{1}t - b(t)}}
{(r_{0} - r)^{3 - a_{1}/\beta}},
\eq
where $(),_{x} = \partial /\partial x$, a prime denotes the ordinary
derivative with respect to the indicated argument, and $a_{1}$ and
$\rho^{out (0)}_{-}$ are two constants, while $A(r)$ and $b(t)$ are
given, respectively, by Eqs.(\ref{2.9}) and (\ref{2.3}). Projecting
$T^{- out} _{\mu\nu}$ onto the PPON frame defined by Eqs.(A.6) and
(A.7), we find that the non-vanishing components are given by
\bqn
\lb{3.4}
 T^{- out} _{(0)(0)} &=& T^{- out} _{(1)(1)} = T^{- out} _{(0)(1)}
=  \frac{\rho^{out (0)}_{-}e^{a_{1}t - b(t)}}
{(r_{0} - r)^{3 - a_{1}/\beta}} \nb\\
&\times& \left\{\frac{E^{2}}{\beta^{2}(r_{0} - r)^{2}} 
- \frac{E[E^{2} - \beta^{2}(r_{0} - r)^{2}]^{1/2}}
{[\beta (r_{0} - r)]^{2}} - \frac{1}{2}\right\}.
\eqn
Clearly, for $b(t) =
\ln[\cosh \beta t], \; \beta t$, we have to choose $a_{1} = \beta$ in order
to have the perturbations be finite initially ($t = - \infty$), while for
$b(t) = - \beta t$, we have to choose $a_{1} = - \beta$, namely,
\bq
\lb{3.5}
 a_{1} = \left\{ \begin{array}{ll}
\beta, & b(t) = \ln[\cosh \beta t], \; \beta t,\\
- \beta, & b(t) = - \beta t. \end{array}\right. 
\eq
Eqs.(\ref{3.4}) and (\ref{3.5}) show that all the components diverge as
$r \rightarrow r_{0}^{-}$ for all the three different choices of $b(t)$,
which indicates that if we take the back-reaction of the null dust fluid into
account, the CEHs appearing on the hypersurface $r = r_{0}$ in the
solutions (\ref{2.10}) and (\ref{2.14}) will be turned into spacetime
singularities, provided that the HK conjecture still holds here.  Since
all the corresponding fourteen scalars constructed from the Riemann
tensor are zero in the present case, the resulted singularities would be
expected to be non-scalar curvature singularities.

In addition to the out-going null dust fluid, if there also exists an
in-going null fluid moving along the null geodesics defined by $l^{\mu}$,
i.e., 
\bq
\lb{3.6}
T^{- in}_{\mu\nu} = \rho^{in}_{-}  l_{\mu} l_{\nu},
\eq
where $l_{\mu}$ is defined by Eq.(\ref{A.11}), then, from the
conservation equations $ T^{- in}_{\mu\nu;\lambda}g^{\nu\lambda} = 0$,
one can show that $\rho^{in}_{-}$ is give by
\bq
\lb{3.7}
 \rho^{in}_{-}  = \frac{\rho^{in (0)}_{-}e^{a_{0}t - b(t)}}
{(r_{0} - r)^{3 + a_{0}/\beta}},
\eq
where $a_{0}$ and $\rho^{in (0)}_{-}$ are other two integration constants.
Then, the   non-vanishing tetrad components of $T^{- in}_{\mu\nu}$ 
are given by
\bqn
\lb{3.8}
 T^{- in}_{(0)(0)} &=& T^{- in}_{(1)(1)} = - T^{- in}_{(0)(1)}
= \frac{\rho^{in (0)}_{-}e^{a_{0}t - b(t)}}
{(r_{0} - r)^{3 + a_{0}/\beta}}\nb\\
&\times& \left\{\frac{E^{2}}{\beta^{2}(r_{0} - r)^{2}} 
+ \frac{E[E^{2} - \beta^{2}(r_{0} - r)^{2}]^{1/2}}
{[\beta (r_{0} - r)]^{2}} - \frac{1}{2}\right\}.
\eqn
Similar to the out-going case, to have the perturbations be finite
initially ($t = - \infty$), we have to choose $a_{0} = a_{1}$, where
$a_{1}$ is given by Eq.(\ref{3.5}). Then, from Eq.(\ref{3.8}) we can see
that these components also diverge. Since now we have
\bq
\lb{3.9}
 T^{-\mu\nu}T^{-}_{\mu\nu} = 2\rho^{out}_{-} \rho^{in}_{-}  
= 2\rho^{out (0)}_{-}\rho^{in (0)}_{-} \frac{e^{2[a_{0}t - b(t)]}}
{(r_{0} - r)^{6}},
\eq
which always diverges as $r \rightarrow r_{0}^{-}$, we can see that the
resulted singularities should be scalar curvature ones when the two null
dust fluids are all present, where
\bq
\lb{3.10}
T^{-}_{\mu\nu} \equiv T^{- out}_{\mu\nu} + T^{- in}_{\mu\nu}.
\eq

Similarly, we can consider test null dust fields in the region $r \ge
r_{0}$, and will obtain the same conclusions. Thus, the above
considerations suggest that all the CEHs appearing in the solutions
(\ref{2.10}) and (\ref{2.14}) are not stable against perturbations for
all the three different choices of $b(t)$ with $b_{0} > 0$.

\section*{IV. PERTURBATIONS NEAR THE CEHs}

\renewcommand{\theequation}{4.\arabic{equation}}
\setcounter{equation}{0}

In this section, let us consider perturbations of null dust fluids near
the CEHs. For the sake of convenience, we shall work with the
coordinates $t$ and $R$, in terms of which the metrics (\ref{2.10}) and
(\ref{2.14}) can be cast in the form
\bq
\lb{4.1}
ds^{2} = e^{- \Omega_{(0)}} (dt^{2} - dR^{2}) - 
e^{- h_{(0)}}\left[e^{\Phi_{(0)}}dz^{2} + e^{- \Phi_{(0)}} d\theta^{2}
\right],
\eq
where
\bqn
\lb{4.2}
\Omega_{(0)} &=& 2 \beta R, \;\;\; h_{(0)} = \beta R - b(t) - \ln C_{0},\nb\\
\Phi_{(0)} &=& - \beta R + b(t) - \ln C_{0},
\eqn
for $r\le r_{0}$, and 
\bqn
\lb{4.3}
\Omega_{(0)} &=& - 2 \beta R, \;\;\; h_{(0)} = - 
\left[\beta R + b(t) + \ln C_{0}\right],\nb\\
\Phi_{(0)} &=& \beta R + b(t) - \ln C_{0},
\eqn
for $r \ge r_{0}$, and the function $b(t)$ is given by Eq.(\ref{2.3}).

As shown in \cite{LW1994}, the null dust fluids given by
\bq
\lb{4.6a}
T_{\mu\nu} = \rho^{in} l_{\mu} l_{\nu} + \rho^{out}  n_{\mu} n_{\nu},
\eq
have contributions only to the metric coefficients $g_{tt}$
and $g_{RR}$. Specifically, if we set 
\bq 
\lb{4.4} 
\left\{\Omega, h, \Phi\right\} =
\left\{\Omega_{(0)} + f(u) + g(v), h_{(0)}, \Phi_{(0)}\right\}, 
\eq 
the metric
\bq
\lb{4.5}
ds^{2} = e^{- \Omega} (dt^{2} - dR^{2}) - 
e^{- h}( e^{\Phi}dz^{2} + e^{- \Phi} d\theta^{2}),
\eq
will satisfy the Einstein field equations $R_{\mu\nu} - g_{\mu\nu} R/2 =
T_{\mu\nu}$, with $\rho^{out}$ and $\rho^{in}$ being given,
respectively, by
\bq
\lb{4.6}
\rho^{out}  = g'(v)h_{,v}, \;\;\;\; \rho^{in}  = f'(u) h_{,u},
\eq
and now
\bqn
\lb{4.7}
l_{\mu} &=& e^{- \Omega/2}(\delta^{t}_{\mu} + \delta^{R}_{\mu}), \;\;\;
n_{\mu} = e^{- \Omega/2}(\delta^{t}_{\mu} - \delta^{R}_{\mu}), \nb\\
u &\equiv& \frac{t + R}{\sqrt{2}}, \;\;\;\;\; 	
v \equiv \frac{t - R}{\sqrt{2}},
\eqn
and $f(u)$ and $g(v)$ are arbitrary functions of their indicated
arguments.  Note that although $T_{\mu\nu}$ now takes the same form as
that considered in the last section, it has fundamental difference: now
it acts as a source of the spacetime. As a result, the back reaction of
it is automatically fully taken into account. When $f(u), g(v)$ and
their first derivatives are very small, the two dust fluids can be
considered as perturbations of the spacetime given  by Eqs.(\ref{4.1}) -
(\ref{4.3}).  In the following, let us consider the three cases $b(t) =
\ln[\cosh(\beta t)], \; + \beta t, \; - \beta t$, separately.

\vspace{1.cm}

\centerline{\bf A. $b(t) = \ln[\cosh(\beta t)]$}

In this case, Eqs.(\ref{4.2}) - (\ref{4.4}) yield
\bqn
\lb{4.8}
\Omega &=& f^{-}(u) + g^{-}(v) + 2 \beta R,\nb\\
h &=& \beta R - \ln[\cosh(\beta t)] - \ln C_{0},\nb\\
\Phi &=& - \beta R + \ln[\cosh(\beta t)] - \ln C_{0},
\eqn
for $r \le r_{0}$, and 
\bqn
\lb{4.9}
\Omega &=& f^{+}(u) + g^{+}(v) - 2 \beta R,\nb\\
h &=& -\left\{\beta R + \ln[\cosh(\beta t)] + \ln C_{0}\right\},\nb\\
\Phi &=&  \beta R + \ln[\cosh(\beta t)] - \ln C_{0},
\eqn
for $r \ge r_{0}$. Substituting Eqs.(\ref{4.8}) and (\ref{4.9}) into
Eq.(\ref{4.6}), we find
\bqn
\lb{4.10}
\rho^{out}_{-}  &=& - \frac{\beta e^{\beta t} {g^{-}}'(v)}
{\sqrt{2}\cosh \beta t}, \;\;\;
\rho^{in}_{-}  = \frac{\beta e^{- \beta t} {f^{-}}'(u)}
{\sqrt{2}\cosh \beta t}, \;(r \le r_{0}),\nb\\
\rho^{out}_{+} &=& \frac{\beta e^{- \beta t} {g^{+}}'(v)}
{\sqrt{2}\cosh \beta t}, \;\;\;
\rho^{in} _{+} = - \frac{\beta e^{\beta t} {f^{+}}'(u)}
{\sqrt{2}\cosh \beta t}, \;(r \ge r_{0}).
\eqn
Note that it is not necessary to take $f^{-}(u)$ and
$g^{-}(v)$ the same forms as $f^{+}(u)$ and $g^{+}(v)$, since now we
consider the perturbations in both sides of the hypersurface $ r =
r_{0}$ independently. However, to have physically reasonable
perturbations, we require
\bqn
\lb{4.11}
{g^{-}}'(v) & < & 0, \;\;\; {g^{+}}'(v) > 0,\nb\\
{f^{-}}'(u) & > & 0, \;\;\; {f^{+}}'(u) < 0,
\eqn
so that $\rho^{out}_{\pm} $ and $\rho^{in}_{\pm} $ are all no negative.

When $f^{\pm}(u),\; g^{\pm}(v)$ and their first derivatives are very
small, the radial time-like geodesics given by Eq.(\ref{A.6}) would be a
very good approximation of the corresponding ones of the metric
(\ref{4.5}). Consequently, the tetrad frames given by Eqs.(\ref{A.6})
and (\ref{A.7}) would serve well as the corresponding PPON of
Eq.(\ref{4.5}). Projecting the EMT onto this frame, we find that the
non-vanishing tetrad components of it are given by
\bqn
\lb{4.12}
 T^{\pm}_{(0)(0)} &=& T^{\pm}_{(1)(1)} =  \frac{1}{2}
 \left\{ D_{+}(A) \rho^{in}_{\pm} + D_{-}(A) \rho^{out}_{\pm}
\right\}, \nb\\
 T^{\pm}_{(0)(1)} &=& T^{\pm}_{(1)(0)} =  \frac{1}{2}
 \left\{ D_{+}(A) \rho^{in}_{\pm} - D_{-}(A) \rho^{out}_{\pm}
\right\},  
\eqn
where $\rho^{in}_{\pm},\; \rho^{out}_{\pm}$ are given by Eqs.(\ref{4.10}), 
and  
\bq
\lb{4.13}
 D_{\pm}(A) = \left[ \frac{E}{e^{A}} \pm \epsilon 
\left(\frac{E^{2}}{e^{2A}} - 1 \right)^{1/2}\right]^{2}, 
\eq
with $A$ being given by Eqs.(\ref{2.9}) and (\ref{2.17}). Clearly, as $r
\rightarrow r_{0}$, these components all diverge. Note that in writing
the above expressions we have used the fact that $f^{\pm}(u)$ and
$g^{\pm}(v)$ are very small to set $exp\{f^{\pm}(u) + g^{\pm}(v)\} = 1$.
This will be also the case for other two cases to be considered below.
Combining Eqs.(\ref{4.10}) with Eqs.(\ref{4.12}) and (\ref{4.13}), we
can see that the perturbations are finite at the initial $t = - \infty$,
but all will focus into a spacetime singularity when they arrive at $r =
r_{0}$. That is, the perturbations turn the CEHs into spacetime
curvature singularities. The nature of the singularity is a scalar one.
This can be seen, for example, from the Kretshmann scalar,
\bqn
\lb{4.15}
 {\cal R} &=& R_{\alpha\beta\gamma\delta} R^{\alpha\beta\gamma} \nb\\
&=&  \left\{ \begin{array}{ll}
- \frac{4 \beta^{2} e^{2[2\beta R + f^{-}(u) + g^{-}(v)]}}
   {\cosh^{2}\beta t}
{f^{-}}'(u){g^{-}}'(v), & r \le r_{0}, \\
           &   \\
  \frac{4 \beta^{2} e^{- 2[2\beta R - f^{+}(u) - g^{+}(v)]}}
   {\cosh^{2}\beta t}
{f^{+}}'(u){g^{+}}'(v), & r \ge r_{0}, \end{array}\right.
\eqn
which diverges like $(\tau_{0} - \tau)^{-1}$, as $r \rightarrow r_{0}$,
as long as ${f^{\pm}}'(u){g^{\pm}}'(v) \not= 0$, as one can show from
Eqs.(\ref{A.9a}) and (\ref{A.9b}).  When only the outgoing null dust fluid
exists, the singularity degenerates to a non-scalar curvature
singularity, as it can be shown that now all the fourteen scalars made
out of the Riemann tensor are zero.  However, in any case the
singularity is strong in the sense that the distortion, which is equal
to the twice integral of the tetrad components of the Riemann tensor,
becomes unbounded, for example,
\bqn
\lb{4.16}
\int  \int R^{\pm}_{(0)(2)(0)(2)}d\tau d\tau &=& 
\pm \frac{\sqrt{2} \beta}{4}
\int \int \frac{1}{\cosh\beta t}\nb\\
& & \times \left[
D_{+}(A)e^{\pm \beta (t - 2R)}{f^{\pm}}'(u) \right.\nb\\
& & \left. - D_{-}(A)e^{\mp \beta (t + 2R)}{g^{\pm}}'(v)\right]
d\tau  d\tau \nb\\          
&\sim&  {g^{\pm}}'(v) \ln(\tau_{0} - \tau),
\eqn
as $r \rightarrow r_{0}$.
It is interesting to note that, although the tetrad components of
$T_{\mu\nu}$ are singular as $r \rightarrow r_{0}$, the scalar
$T^{\mu\nu}T_{\mu\nu}$ does not. In fact, from Eq.(\ref{4.12}) it can be
shown that
\bqn
\lb{4.17}
T^{\pm}_{\mu\nu} T^{\pm \mu\nu} &=& T^{\pm}_{(a)(b)}T^{\pm(a)(b)}
= 2 \rho^{in}_{\pm}\rho^{out}_{\pm} = \nb\\
&=& - \frac{\beta^{2}{f^{\pm}}'(u){g^{\pm}}'(v)}{2 \cosh^{2}\beta t}
\sim (\tau_{0} - \tau),
\eqn
as $r \rightarrow r_{0}$. Thus, the formation of the spacetime singularity 
is mainly due to the focus of the corresponding gravitational fields.
This is different from what we can get from Eq.(\ref{3.9}) 
for the test particles.

\vspace{1.cm}

\centerline{\bf B. $b(t) = + \beta t$}

When $b(t) = + \beta t$, from Eqs.(\ref{4.2}) - (\ref{4.4}) we find that
\bqn
\lb{4.18}
\Omega &=& f^{-}(u) + g^{-}(v) + 2 \beta R,\nb\\
h &=& - \beta (t - R) - \ln C_{0},\nb\\
\Phi &=& \beta (t - R)  - \ln C_{0},
\eqn
for $r \le r_{0}$, and 
\bqn
\lb{4.19}
\Omega &=& f^{+}(u) + g^{+}(v) - 2 \beta R,\nb\\
h &=& - \beta (t + R) - \ln C_{0},\nb\\
\Phi &=&  \beta (t + R) - \ln C_{0},
\eqn
for $r \ge r_{0}$. Substituting Eqs.(\ref{4.18}) and (\ref{4.19}) into
Eq.(\ref{4.6}), we find that
\bqn
\lb{4.20}
\rho^{out}_{-}  &=& - \sqrt{2}\beta {g^{-}}'(v), \;\;\;
\rho^{in}_{-}  = 0, \;(r \le r_{0})\nb\\
\rho^{out}_{+} &=& 0, \;\;\;
\rho^{in} _{+} = - \sqrt{2}\beta {f^{+}}'(u), \;(r \ge r_{0}).
\eqn
The above expressions show that in the region $r \le r_{0}$ now there
exists only outgoing dust cloud, while in the region $r \ge r_{0}$ only
ingoing. With the same arguments as those given in the last subsection,
we take the tetrad frames given by Eqs.(\ref{A.6}) and (\ref{A.7}) as a
good approximation to the corresponding PPON of Eq.(\ref{4.5}), when
$f^{\pm}(u), \; g^{\pm}(v)$, and their first derivatives are very small.
Then,  projecting the EMT onto this frame, we find that the
non-vanishing tetrad components of it are given by
\bqn
\lb{4.23}
 T^{-}_{(0)(0)} &=& T^{-}_{(1)(1)} = -  
 T^{-}_{(0)(1)}  = - \frac{\beta}{\sqrt{2}}
   D_{-}(A) {g^{-}}'(v), \; (r \le r_{0}),\nb\\ 
  T^{+}_{(0)(0)} &=& T^{+}_{(1)(1)} =   
 T^{+}_{(0)(1)}  = - \frac{\beta}{\sqrt{2}}
   D_{+}(A) {f^{+}}'(u), \; (r \ge r_{0}),
\eqn
with $A$ being given by Eqs.(\ref{2.9}) and (\ref{2.17}), and $D_{\pm}$
are defined by Eq.(\ref{4.13}). Clearly, as $r \rightarrow r_{0}$, these
components all diverge, although at the initial $t = -\infty, r \not=
r_{0}$ they are finite. That is, the perturbations, similar to the
last subcase, turn the CEHs into spacetime curvature singularities. The
nature of the singularity is a non-scalar one, as one can show that now
all the fourteen scalars built from the Riemann tensor are zero.
However, the singularity is strong in the sense that the distortion
diverges like $\ln (\tau_{0} - \tau)$ as $r \rightarrow r_{0}$, as we 
can see from the following integrations,
\bqn
\lb{4.25}
\int  \int R^{-}_{(0)(2)(0)(2)}d\tau d\tau &=& 
 \frac{1}{\sqrt{2}\beta}
\int \int \frac{{g^{-}}'(v)}{(r_{0} - r)^{2}}D_{-}(A)
d\tau  d\tau \nb\\          
&\sim&  \ln(\tau_{0} - \tau),\nb\\
\int  \int R^{+}_{(0)(2)(0)(2)}d\tau d\tau &=& 
 \frac{1}{\sqrt{2}\beta}
\int \int \frac{{f^{+}}'(u)}{(r - r_{0})^{2}}D_{+}(A)
d\tau  d\tau \nb\\          
&\sim&  \ln(\tau_{0} - \tau).
\eqn

\vspace{1.cm}

\centerline{\bf C. $b(t) = - \beta t$}

When $b(t) = - \beta t$, Eqs.(\ref{4.2}) - (\ref{4.4}) yield
\bqn
\lb{4.26}
\Omega &=& f^{-}(u) + g^{-}(v) + 2 \beta R,\nb\\
h &=& \beta (t + R) - \ln C_{0},\nb\\
\Phi &=& - \beta (t + R)  - \ln C_{0},
\eqn
for $r \le r_{0}$, and 
\bqn
\lb{4.27}
\Omega &=& f^{+}(u) + g^{+}(v) - 2 \beta R,\nb\\
h &=& \beta (t - R) - \ln C_{0},\nb\\
\Phi &=& - \beta (t - R) - \ln C_{0},
\eqn
for $r \ge r_{0}$. Substituting Eqs.(\ref{4.26}) and (\ref{4.27}) into
Eq.(\ref{4.6}), we find
\bqn
\lb{4.28}
\rho^{out}_{-}  &=& 0, \;\;\;
\rho^{in}_{-}  = \sqrt{2}\beta {f^{-}}'(u), \;(r \le r_{0}), \nb\\
 \rho^{out}_{+} &=& \sqrt{2}\beta {g^{+}}'(v), \;\;\;
\rho^{in} _{+} = 0, \;(r \ge r_{0}).
\eqn
Thus, in the present case in the region $r \le
r_{0}$ there exists only ingoing dust cloud, while in the region $r \ge
r_{0}$ only outgoing. Therefore, now the dust clouds cannot be considered
as perturbations, but rather than as emission of null fluids from the CEHs.  To
study the stability of the CEHs in this case we have to consider other
kinds of perturbations. However, the following considerations indicate
that they may be not stable, too.  Projecting the EMT onto the frame given
by Eqs.(\ref{A.6}) and (\ref{A.7}), we find that
\bqn
\lb{4.30}
 T^{-}_{(0)(0)} &=& T^{-}_{(1)(1)} = 
 T^{-}_{(0)(1)}  = \frac{\beta}{\sqrt{2}}
   D_{+}(A) {f^{-}}'(u), \; (r \le r_{0}), \nb\\
 T^{+}_{(0)(0)} &=& T^{+}_{(1)(1)} = -  
 T^{+}_{(0)(1)}  = \frac{\beta}{\sqrt{2}}
   D_{+}(A) {g^{+}}'(v), \; (r \ge r_{0}).
\eqn
From Eq.(\ref{4.13}) and the above expressions we can see that the
back reaction of the emission also turns the CEHs into spacetime
singularities. Similar to the last subcase, the nature of the
singularity is a non-scalar one but strong, as the distortion also
diverges as $r \rightarrow r _{0}$,  
\bqn
\lb{4.32}
\int  \int R^{-}_{(0)(2)(0)(2)}d\tau d\tau &=& 
 - \frac{1}{\sqrt{2}\beta}
\int \int \frac{{f^{-}}'(u)}{(r_{0} - r)^{2}}D_{+}(A)
d\tau  d\tau \nb\\          
&\sim&  \ln(\tau_{0} - \tau),\nb\\
\int  \int R^{+}_{(0)(2)(0)(2)}d\tau d\tau &=& 
 - \frac{1}{\sqrt{2}\beta}
\int \int \frac{{g^{+}}'(v)}{(r - r_{0})^{2}}D_{-}(A)
d\tau  d\tau \nb\\          
&\sim&  \ln(\tau_{0} - \tau).
\eqn
Thus, for the perturbations that have non-vanishing components along the
ingoing null geodesics defined by $l_{\mu}$ in the region $r \le r_{0}$,
or for the perturbations that have non-vanishing components along the
outgoing null geodesics defined by $n_{\mu}$ in the region $r \ge
r_{0}$, we would expect that the CEHs will be turned into spacetime
singularities.

\section*{V. CONCLUDING REMARKS }

In this paper, we have considered the stability of the CEHs for a class
of non-static global cosmic strings found recently by Gregory
\cite{Gregory1996}, and found that they are not stable against
perturbations. In particular, the back reaction of null dust fluids will
turn them into spacetime singularities. Thus resulted
singularities are strong in the sense that the distortion of test
particles diverges when these singular hypersurfaces are approaching.

Recently, we have shown that the CEHs of topological domain walls are
also not stable against massless scalar field \cite{Wang1992} and null
dust fluids \cite{Wang1995}. Thus, a natural question is that: Are all
the cosmological event horizons not stable? If some are but others not,
what are the criteria for them? It was exactly this consideration that
motivated us to study the test particles in Sec. III, at the aim of
generalizing the HK conjecture to the study of the stability of the
CEHs.  Comparing the results obtained in Sec. III with the ones obtained
in Sec. IV for real perturbations, we can see that the HK conjecture can
be used directly to the study of the stability of CEHs (as far as the
examples considered in this paper are concerned), except for the case
where $b(t) = \ln[\cosh(\beta t)]$. In the latter case, although the
study of the test particles gives a correct prediction for the nature of
the resulted singularities, but the quantity $T_{\mu\nu} T^{\mu\nu}$ for
test particle diverges, while for the real perturbations it does not. As
a matter of fact, the divergence of the Kretschmann scalar is mainly due
to the non-linear interaction of gravitational fields for the real
perturbations. Clearly, to properly form the conjecture, more examples
need to be considered.

Finally, we would like to note that although the CEHs found by Gregory
are not stable, and after the back reaction of perturbations is taken
into account, they will be turned into spacetime singularities, the hope
that the time-dependence of the spacetimes for global cosmic strings may
remove the singular nature of the corresponding static ones has not been
shown completely negative. As we mentioned in the introduction, the
class of spacetimes considered by Gregory is not the most general
spacetimes for non-static global cosmic strings. In some cases, one may
expect that the gravitational and particle radiation is so strong that
the gravitational field of a global string may be well dilated before
any spacetime singularity or event horizon is formed. Spacetimes with
cylindrical symmetry are quite different from those with spherical
symmetry. As a matter of fact, in the former case gravitational
radiation generally always exists.


\section*{APPENDIX}

\renewcommand{\theequation}{A.\arabic{equation}}
\setcounter{equation}{0}

In this appendix, we shall briefly review some main properties of the
spacetimes given by
\bq
\lb{A.1}
ds^{2} = e^{2A(r)}dt^{2} - dr^{2} - e^{2[A(r) + b(t)]}dz^{2} 
- C^{2}(r)d\theta^{2},
\eq
where $A, b$ and $C$ are arbitrary functions of their indicated arguments.

The corresponding Lagrangian of the radial geodesics is given by
\bq
\lb{A.2}
{\cal L} = \frac{1}{2} \left(\frac{ds}{d\tau}\right)^{2}
= \frac{1}{2} \left[e^{2A(r)}\dot{t}^{2} - \dot{r}^{2} \right],
\eq
where an overdot denotes the ordinary derivative with respect to the
parameter $\tau$ of the geodesics. For time-like geodesics, $\tau$ can
be identified as the proper time of the test particles. Since ${\cal L}$
is independent of $t$, the quantity $E$ given by
\bq
\lb{A.3}
E \equiv \frac{\partial {\cal L}}{\partial \dot{t}} = e^{2A(r)} \dot{t},
\eq
is an integration constant, which represents the total energy of the
test particles for time-like geodesics. Substituting Eq.(\ref{A.3}) into
Eq.(\ref{A.2}), we find
\bq
\lb{A.4}
\dot{r} = \epsilon \left(\frac{E^{2}}{e^{2A(r)}} - 2 
{\cal L}\right)^{1/2},\; (\epsilon = \pm 1),
\eq
where $\epsilon = + 1$ corresponds to the out-going geodesics, while
$\epsilon = - 1$ to the in-going geodesics. By properly choosing the
parameter $\tau$, we can always make ${2\cal{L}} = + 1, 0, - 1$,
respectively, for time-like, null, and space-like geodesics. In the
following, we shall consider only time-like geodesics. Then, the motion
of the test particles has the first integral
\bq
\lb{A.5}
\dot{t} = \frac{E}{e^{2A(r)}}, \;\;\;   
\dot{r} = \epsilon \left(\frac{E^{2}}{e^{2A(r)}} - 1\right)^{1/2}.
\eq
Denoting the tangent vector to the geodesics by $\lambda^{\mu}_{(0)}$,
\bq
\lb{A.6}
\lambda^{\mu}_{(0)} \equiv \frac{dx^{\mu}}{d\tau} 
= \dot{t}\delta_{t}^{\mu} + \dot{r}\delta_{r}^{\mu}
= \dot{t}\delta_{t}^{\mu} + e^{- a(R)}\dot{r}\delta_{R}^{\mu},
\eq
we can construct other three orthogonal space-like vectors
\bqn
\lb{A.7}
\lambda^{\mu}_{(1)} &=& e^{-A(r)} \dot{r}\delta_{t}^{\mu} + 
e^{A(r)}\dot{t}\delta_{r}^{\mu}
= e^{- a(R)} \dot{r}\delta_{t}^{\mu} + 
 \dot{t}\delta_{R}^{\mu},\nb\\
\lambda^{\mu}_{(2)} &=& e^{-[A(r) + b(t)]} \delta_{z}^{\mu},\;\;\;\;
\lambda^{\mu}_{(3)} = C^{-1}(r) \delta_{\theta}^{\mu}. 
\eqn
Then, it can be shown that
\bq
\lb{A.8}
\lambda^{\mu}_{(i)} \lambda_{(j) \mu} = \eta_{i j},\;\;\;
\lambda^{\mu}_{(i); \nu} \lambda_{(0)}^{\nu} = 0, \;
(i, j = 0, 1, 2, 3),
\eq
where $\eta_{i j}$ is the Minkowski metric. The above equations show
that the four unit vector $\lambda^{\mu}_{(i)}$ form a freely-falling
frame or parallel-propagated orthogonal frame (PPON) along the time-like
geodesics.

For the particular solutions of $A(r)$ given by Eqs.(\ref{2.9}) and
(\ref{2.17}), Eq.(\ref{A.5}) has the following integration,
\bqn
\lb{A.9a}
e^{-2\beta R} &=& \beta^{2} (r_{0} - r)^{2} = 
\beta^{2}(\tau_{0}^{2} - \tau^{2}),\nb\\
e^{2\beta t} &=& \frac{\tau_{0} + \tau}{\tau_{0} - \tau},\; (r \le r_{0}),
\eqn
for $r \le r_{0}$, and 
\bqn
\lb{A.9b}
e^{2\beta R} &=& \beta^{2} (r - r_{0})^{2} = 
\beta^{2}(\tau_{0}^{2} - \tau^{2}),\nb\\
e^{2\beta t} &=& \frac{\tau_{0} + \tau}{\tau_{0} - \tau},\; (r \ge r_{0}),
\eqn
for $r \ge r_{0}$, where $\tau_{0}$ is chosen such that when $r
\rightarrow r_{0}$, we have $\tau \rightarrow \tau_{0}$.

On the other hand, the corresponding Kretschmann scalar to the metric
(\ref{A.1}) is given by
\bqn
\lb{A.10}
{\cal R} &\equiv& R_{\alpha\beta\gamma\lambda}R^{\alpha\beta\gamma\lambda}
=  4\left\{\left(\frac{C''}{C}\right)^{2} + 
2\left(\frac{A'C'}{C}\right)^{2}\right\}\nb\\
&& + 4 \left\{ 2(A'' + A'^{2})^{2} + 
[A'^{2} - (b'' + b'^{2})e^{-2A}]^{2}\right\}.
\eqn

Choosing a null tetrad frame, on the other hand, as 
\bqn
\lb{A.11}
l_{\mu} &=& \frac{1}{\sqrt{2}}\left(e^{A(r)}\delta^{t}_{\mu} 
+ \delta^{r}_{\mu}\right) = \frac{e^{a(R)}}{\sqrt{2}}
(\delta^{t}_{\mu} + \delta^{R}_{\mu}),\nb\\
n_{\mu} &=& \frac{1}{\sqrt{2}}\left(e^{A(r)}\delta^{t}_{\mu} 
- \delta^{r}_{\mu}\right) = \frac{e^{a(R)}}{\sqrt{2}}
(\delta^{t}_{\mu} - \delta^{R}_{\mu}),\nb\\
m_{\mu} &=& \frac{1}{\sqrt{2}}\left[e^{A(r) + b(t)}\delta^{z}_{\mu} 
+ i C(r) \delta^{\theta}_{\mu}\right],\nb\\
\bar{m}_{\mu} &=& \frac{1}{\sqrt{2}}\left[e^{A(r) + b(t)}\delta^{z}_{\mu} 
- i C(r) \delta^{\theta}_{\mu}\right],
\eqn
we find that the non-vanishing Weyl scalars are given by
\bqn
\lb{A.12}
\Psi_{2} &=& - \frac{1}{2}C_{\mu\nu\lambda\delta}
\left[l^{\mu}n^{\nu}l^{\lambda}n^{\delta} -
l^{\mu}n^{\nu}{m}^{\lambda}\bar{m}^{\delta}\right]\nb\\
&=& \frac{1}{12}\left\{ -\frac{C''}{C} + A'' + \frac{C'}{C}A' 
+ (b'' + b'^{2})e^{-2A}\right\},\nb\\
\Psi_{0} &=& - C_{\mu\nu\lambda\delta}l^{\mu}m^{\nu}
l^{\lambda}m^{\delta} = - 3\Psi_{2}, \nb\\
\Psi_{4} &=& - C_{\mu\nu\lambda\delta}n^{\mu}\bar{m}^{\nu}
n^{\lambda}\bar{m}^{\delta} = - 3\Psi_{2}.
\eqn
Thus, we have
\bq
\lb{A.13}
\Psi_{0}\Psi_{4} = 9 \Psi_{2}^{2}.
\eq
Then, according to the theorem given in \cite{SX1986}, we find that the
metric (\ref{A.1}) is always Petrov type D, except for the degenerate
case where $\Psi_{2} = 0$, which is Petrov type O.

Finally, we would like to note that $l_{\mu}\; (n_{\mu})$ defines an
ingoing (outgoing) radial null geodesic congruence \cite{LW1994}.


\section*{Acknowledgment}

The financial assistance from CNPq (AW), and the ones from  CLAF-CNPq
(JACN) and UMSA (JACN), are gratefully acknowledged.


\end{document}